\documentclass[preprint2]{aastex}

\newcommand{\tfrac}[2]{\textstyle\frac{#1}{#2}\displaystyle}

\shorttitle{Galaxy cluster gas profile}
\shortauthors{Frederiksen et al.}

\begin{document}

\title{Determining all gas properties in galaxy clusters from the dark matter distribution alone}


\author{Teddy F. Frederiksen$^\dag$, Steen H. Hansen$^\dag$, Ole Host$^\dag$, Marco Roncadelli$^\ddag$}
\affil{$^\dag$ Dark Cosmology Centre, Niels Bohr Institute, University of Copenhagen,\\
Juliane Maries Vej 30, 2100 Copenhagen, Denmark}
\affil{$^\ddag$  INFN, Sezione di Pavia, Via A. Bassi 6, 27100 Pavia, Italy}




\begin{abstract}
We demonstrate that all properties of the hot X-ray emitting gas in galaxy clusters are completely 
determined by the underlying dark matter (DM) structure. Apart from the { standard conditions of spherical symmetry and hydrostatic equilibrium 
for the gas, our proof is based on the Jeans equation for the DM and} two simple relations which 
have recently emerged from numerical simulations: the equality of the gas and DM 
temperatures, and the almost linear relation between the DM velocity anisotropy profile 
and its density slope. For DM distributions described by the NFW or the Sersic profiles, the resulting 
gas density profile, the gas-to-total-mass ratio profile, and the entropy profile are all in good agreement 
with X-ray observations. All these profiles are derived using zero free parameters.
Our result allows us to predict the X-ray luminosity profile of a cluster in terms 
of its DM content alone. As a consequence, a new strategy becomes available to constrain the DM 
morphology in galaxy clusters from X-ray observations.  Our results can also be used as a practical tool for 
creating initial conditions for realistic cosmological structures to be used in numerical simulations.

\end{abstract}


\keywords{dark matter, galaxies: clusters: general, X-rays: galaxies: clusters}


\section{Introduction}

Galaxy clusters are the largest equilibrated structures in the Universe, consisting mainly
of dark matter (DM) and hot ionized gas in hydrostatic equilibrium in the overall potential well. 
Observations of this X-ray emitting gas allow for an accurate determination of the properties of the 
dominating DM structure, which can then be compared with the results of numerical N-body 
simulations. 

More specifically, the strategy can be outlined as follows. In the first place, the 
equation of hydrostatic equilibrium can be used to infer the DM density profile from X-ray data 
\citep{fabricant1980}. Application of this technique in conjunction with present-day observations 
\citep{2006MNRAS.368..518V,pointe} yields density profiles which are in excellent agreement 
with those emerging from numerical simulations of structure formation \citep{nfw,moore,diemand,
stadel2008,navarro2008}. Moreover, using a very simple connection between the gas and DM temperatures which has been confirmed by numerical simulations, the equation of hydrostatic equilibrium can be combined with 
the Jeans equation for the DM to derive both the DM radial velocity dispersion and the velocity anisotropy 
profile \citep{hansenpiff,host2009}. Again, the resulting profiles turn out to be in excellent agreement with numerical simulations \citep{colelacey,carlberg}.

Measurements of the gas temperature profile have demonstrated its virtually universal properties
\citep{2002ApJ...567..163D,2004A&A...413..415K,2005ApJ...628..655V, pointe,2006MNRAS.368..518V}.
This universality appears surprising since the temperature profile should encode information about the 
violent gra\-vi\-tational processes taking place during the cluster formation, as well as any additional energy 
input, e.g.~from a central heat engine, and these processes are expected to differ significantly from structure to structure. The gas density profile also exhibits a roughly universal behaviour, which has allowed observers to fit 
remarkably simple forms, e.g.~a beta-profile \citep{1976A&A....49..137C,1986RvMP...58....1S}, to the data. 

A natural question arises as to whether the properties of the hot X-ray emitting gas in galaxy clusters can be 
{\it predicted} from first principles starting from the cluster DM distribution alone. 

{ Attempts in that direction have been pioneered by  \citep{masasu, susama}. Our goal is to go a step further along this direction, relaying upon  information on the internal cluster dynamics which has become available only quite recently.

More specifically}, we will show how the gas density profile can be obtained directly from the underlying DM profile, by 
combining the equation of hydrostatic equilibrium for the gas and the Jeans equation for the DM { under the standard assumption of spherical symmetry}. Besides 
the above-mentioned relation between the gas and DM temperatures, our derivation rests upon a very 
simple connection between the DM velocity anisotropy and the slope of its density profile, which has recently 
emerged in numerical simulations \citep{hansenmoore,hansenstadel}. We thereby demonstrate that the gas 
density profile is completely determined once the gravitationally dominant DM density profile is given. Since the gas 
temperature profile is also known, it turns out that the DM distribution dictates all the gas 
properties {\it uniquely}. Besides conceptually relevant in itself, this fact allows to predict the X-ray luminosity 
profile of a cluster in terms of its DM content alone { by a method similar to the one put forward by \cite{masasu, susama}}. So, a new strategy becomes available to constrain the DM 
morphology in galaxy clusters from X-ray observations. Moreover, our findings can be employed as a practical tool 
for creating initial conditions for realistic cosmological structures to be used in numerical simulations.

\section{Background}

We start by recalling some basic information which will be instrumental for our analysis. We restrict our attention throughout to 
regular clusters, which are supposed to be spherically symmetric and relaxed. The condition of hydrostatic equilibrium 
for the X-ray emitting gas can be written as
\begin{equation}
\label{a1}
\frac{k_B T_g}{\mu m_p} \, \left(\frac{\mathrm{d} \ln \rho_g}{\mathrm{d} \ln r} + \frac{\mathrm{d} \ln T_g}{\mathrm{d} \ln r} \right) 
+ \frac{G M_{\mathrm{tot}}(r)}{r} = 0~,
\end{equation}
where $\rho_g (r)$ and $T_g (r)$ are the gas density and temperature profiles, respectively, $\mu \simeq 0.61$ is the mean 
molecular weight for the intracluster gas,  $m_p$ is the proton mass and $M_{\mathrm{tot}}(r)$ represents the total mass inside 
radius $r$. Two conditions have to be satisfied in order for Eq.~(\ref{a1}) to hold. First, it should be applied to a region considerably 
larger than the gas mean free path, so that local thermodynamic equilibrium is established. Second, the cooling time in that region 
should be larger than the age of the cluster, so that no bulk motion occurs. The latter condition is generally met outside the central 
region, where the presence of a cooling flow often requires Eq.~(\ref{a1}) to be replaced by the Euler equation (with the velocity term 
playing a nonnegligible role). Because of collisional relaxation, the gas velocity distribution is isotropic and its temperature 
can be expressed in terms of the one-dimensional velocity dispersion ${\sigma_g^2}$ as 
\begin{equation}
\label{a2}
T_g = \frac{\mu m_p {\sigma}^2_g}{k_B}~. 
\end{equation}

Assuming complete spherical symmetry for the DM distribution, the two tangential components of the DM velocity 
dispersion, denoted by ${\sigma}_t^2$, are necessarily equal,
but they are generally allowed to differ from the radial 
component ${\sigma}_r^2$, since DM is supposed to be collisionless. It is usual to quantify the DM velocity anisotropy by 
\begin{equation}
\label{a2A}
\beta \equiv 1 - \frac{{\sigma}^2_t}{{\sigma}^2_r} 
\end{equation}
and we find it convenient to introduce the mean DM one-dimensional velocity dispersion $\sigma_{\mathrm{DM}}^2$ as 
\begin{equation}
\label{a3}
\sigma^2_{\mathrm{DM}} \equiv \frac{1}{3} \left( {\sigma}^2_r + 2 {\sigma}^2_t \right) = \left(1 - \frac{2}{3}  \beta \right) {\sigma}^2_r~. 
\end{equation}
Moreover, in analogy with the case of a gas, we also define the DM temperature as \cite{hansenpiff}
\begin{equation}
\label{a4}
T_{\mathrm{DM}} \equiv \frac{\mu m_p {\sigma^2_{\mathrm{DM}}}}{k_B} = \frac{\mu m_p}{k_B} \left(1 - \frac{2}{3}  \beta \right) {\sigma}^2_r~. 
\end{equation}
Of course, the collisionless nature of DM prevents any definition of temperature in the thermodynamic sense and in fact 
$T_{\mathrm{DM}}$ is simply meant to quantify the average velocity dispersion over the three spatial directions. 
Any completely spherically symmetric and relaxed DM configuration obeys the Jeans equation
\begin{equation}
\label{a5}
\sigma^2_r \, \left(\frac{\mathrm{d} \ln \rho_{\mathrm{DM}}}{\mathrm{d} \ln r} + \frac{\mathrm{d} \ln \sigma^2_r}{\mathrm{d} \ln r} 
+ 2 \beta \right) + \frac{G M_{\mathrm{tot}}(r)}{r} = 0~,
\end{equation}
where $\rho_{\mathrm{DM}}(r)$ denotes the DM density profile \citep{binneytremaine}.

\section{The temperature profile}

Early studies of the X-ray emission from regular clusters were based on the assumption of an {\it isothermal} gas distribution, simply 
because the {\it Einstein} observatory and ROSAT were unable to determine the cluster temperature profiles. The observed X-ray emission 
is produced by thermal bremsstrahlung \citep{1986RvMP...58....1S}, so for $T_g = \mathrm{const.}$ it follows that $\rho_g (r)$ 
is proportional to the square root of the deprojected X-ray surface brightness. In such a situation, a good fit to the data was provided by the 
beta-model \citep{1976A&A....49..137C,1986RvMP...58....1S}
\begin{equation}
\label{a6}
\rho_g (r) = \frac{\rho_g (0)}{\left[1 + \left(\frac{r}{a_X} \right)^2 \right]^{3 {\beta_{\mathrm{fit}}}/2}}~,
\end{equation}
where $a_X < 0.5 \, {\mathrm{Mpc}}$ is the X-ray core radius. Note that $\beta_\mathrm{fit}$ has nothing to do with the DM velocity anisotropy. 
Typically, most of the emission comes from the region $r > 0.5 \, 
{\mathrm{Mpc}}$, and so Eq.~(\ref{a6}) can be approximated by the power-law
\begin{equation}
\label{a7}
\rho_g (r) \simeq \rho_g (0) \left(\frac{r}{a_X} \right)^{- 3 {\beta_{{\mathrm{fit}}}}}~.
\end{equation}
Now, by inserting Eq.~(\ref{a7})  and $T_g = \mathrm{const}.$ into Eq.~(\ref{a1}), we find
\begin{equation}
\label{a8}
M_{\mathrm{tot}}(r) = \left( \frac{3 {\beta}_{\mathrm{fit}} {\sigma}^2_g}{G} \right) r~,
\end{equation}
where Eq.~(\ref{a2}) has been used. As is well known, under the assumption of isotropic velocity distribution ($\beta = 0$), a mass 
profile of the form $M(r) \propto r$ describes a singular isothermal sphere (SIS) model in which the velocity dispersion is everywhere 
constant \citep{binneytremaine}. Denoting by $\sigma$ the one-dimensional velocity dispersion, we explicitly have 
$M(r) = \left( 2 \sigma^2 /G \right) r$. Owing to the fact that the leading contribution to $M_{\mathrm{tot}}(r)$ comes from DM, 
it follows that $M_{\mathrm{tot}}(r) \simeq M_{\mathrm{DM}}(r)$. As a consequence, Eq.~(\ref{a8}) can be rewritten as
\begin{equation}
\label{a9}
M_{\mathrm{DM}}(r) \simeq \left( \frac{2 {\sigma^2_{\mathrm{DM}}}}{G} \right) r
\end{equation}
and the comparison of Eqs. (\ref{a8}) and (\ref{a9}) entails in turn
\begin{equation}
\label{a10}
{\sigma^2_{\mathrm{DM}}} \simeq 1.5 \, {\beta}_{\mathrm{fit}} \, {\sigma}^2_g~.
\end{equation}
Observations performed with the {\it Einstein} observatory and ROSAT yield $0.5 < {\beta}_{\mathrm{fit}} < 0.9$ with a median 
${\beta}_{\mathrm{fit}} \simeq 0.67$ \citep{BL1994}. Thus, on average we get 
\begin{equation}
\label{a11}
{\sigma^2_{\mathrm{DM}}} \simeq {\sigma}^2_g~,
\end{equation}
which implies
\begin{equation}
\label{a12}
T_{\mathrm{DM}} \simeq T_g~,
\end{equation}
thanks to Eqs.~(\ref{a2}) and (\ref{a4}).

Only with the advent of the ASCA and {\it Beppo}-SAX satellites did it become possible to measure the cluster temperature profiles, 
which turned out to be described by a polytropic gas distribution to first approximation. Higher-quality data are currently provided 
by {\it Chandra} and {\it XMM-Newton} satellites, which have shown that the gas temperature profiles possess a very simple and 
nearly universal behaviour (see \cite{vikhlinin} for a thorough discussion). Basically, it increases rapidly from a small (possibly 
non-zero) value in the centre, to a maximum at a radius about $0.1\, r_{180}$, and then declines slowly by a factor 
of $2-3$ at  $(0.6-0.8) \,r_{180}$. Here, $r_{180}$ is defined as the radius within which the mean total density is 180 times 
the critical density at the redshift of the cluster. The necessary X-ray background subtraction makes it very difficult to accurately 
measure the temperature further out.

As mentioned above, our main goal is the determination of the gas density profile ${\rho}_g(r)$ once a specific dark matter distribution 
$M_{\mathrm{DM}}(r)$ is given. Supposing as before that $M_{\mathrm{tot}}(r) \simeq M_{\mathrm{DM}}(r)$, it is evident that 
${\rho}_g(r)$ follows from Eq.~(\ref{a1}) provided that ${T}_g(r)$ is specified. Previous studies \citep{masasu, 
susama} accomplished this task by assuming
\begin{equation}
\label{a13A}
{T}_g (r) \simeq \frac{G \mu m_p M_{\mathrm{DM}}(r)}{3 k_B r}~,
\end{equation}
which was suggested to formalize the condition that the gas temperature is close to the virial temperature of the DM. However, the virial 
theorem is a global relation that characterizes a cluster as a whole -- it just arises by integrating the Jeans equation over the system -- 
and so { its validity is open to criticism. Of course, in the lack of any other information about the gas temperature this was the only viable possibility. However, recent progress allows for a considerable improvement on this point.} 

As a matter of fact, this stumbling block can be side-stepped in a remarkably simple fashion. Because of the equivalence principle, the velocity 
of a test particle in an external gravitational field is independent of the particle mass. This circumstance leads to the guess
\begin{equation}
\label{a13}
T_{\mathrm{DM}} (r) = \kappa  \, T_g (r)~.
\end{equation}
This relation was tested against numerical simulations \citep{host2009}, which demonstrated its validity with $\kappa =1$ to a 
very good approximation. These numerical simulations \citep{kay,springel,valdarnini} are reliable only on scales greater than
$\sim 0.1 \,r_{2500}$, while the best X-ray observations are sensitive to a radius which is almost a factor 3 smaller. It is therefore 
possible that heating or cooling may shift $\kappa$ away from unity in the very centre. Hence, outside that region $\kappa=1$ is expected. 
Actually, a look back at Eq.~(\ref{a12}) confirms the remarkable fact that 
$\kappa=1$ holds regardless of the actual shape of the DM velocity anisotropy profile $\beta (r)$. As we shall see, starting from a specific 
underlying DM density profile $\rho_{\mathrm{DM}}(r)$, one can evaluate $T_{\mathrm{DM}}(r)$ and then get the gas temperature profile $T_g(r)$ uniquely.

Before closing this section, a remark is in order. Observations show that some clusters lack a central cooling flow. In such a situation, 
hydrostatic equilibrium is expected to hold all the way down to the centre. Actually, for typical central values of the electron number 
density $n_e \simeq1 {\mathrm{cm}}^{-3}$ and temperature $T\simeq10^8 \, {\mathrm{K}} \simeq 8.5 \, {\mathrm{keV}}$ 
\citep{1986RvMP...58....1S}, the scattering time turns out to be $t_{\mathrm{scat}} \sim 10^2 \mathrm{yr}$, which is much smaller than 
the corresponding gas cooling time $t_{\mathrm{cool}} \sim 10^{7} \mathrm{yr}$, so that local hydrostatic equilibrium is indeed fulfilled outside a central spherical region of radius $\sim 1 \, {\rm pc}$. Assuming further that the gas temperature is roughly constant in the inner cluster region, the gas density profile cannot be cuspy as long as $M_{\mathrm{DM}}(r) \propto r^a$ with $a > 1$ for $r \to 0$. This is at odds with blind extrapolations of fitting formulae for the temperature and density such as those used in \cite{vikhlinin}.

\section{The density profile}

We now proceed to the actual derivation of the gas density profile ${\rho_g} (r)$ from the properties of the dominating DM distribution.

As a preliminary step, we notice that Eqs.~(\ref{a1}) and (\ref{a5}) can be trivially combined to yield
\begin{eqnarray}
\label{a15}		
		&&\frac{k_B T_g}{\mu m_p} \, \left(
			\frac{\mathrm{d} \ln \rho_g}{\mathrm{d} \ln r}
			+ \frac{\mathrm{d} \ln T_g}{\mathrm{d} \ln r}
		\right)
		\nonumber \\
		&& =
		\sigma^2_r \, \left(
			\frac{\mathrm{d} \ln \rho_{\mathrm{DM}}}{\mathrm{d} \ln r}
			+ \frac{\mathrm{d} \ln \sigma^2_r}{\mathrm{d} \ln r}
			+ 2 \beta
		\right) \, .
\end{eqnarray}
Owing to Eqs.~(\ref{a4}) and (\ref{a13}) with $\kappa = 1$, straightforward manipulations permit to recast Eq.~(\ref{a15}) into the form 
\begin{eqnarray} 
\label{a16}			
\gamma_g = \frac{ 1 }{ 1 - \frac{2}{3} \beta} \left(\gamma _{\mathrm{DM}} + 2 \beta
			+ \tfrac{2}{3} \beta \frac{\mathrm{d} \ln \sigma^2_r}{\mathrm{d} \ln r}
			+ \frac{2}{3} \frac{\mathrm{d} \beta}{\mathrm{d} \ln r} \right)~,
\end{eqnarray}
where we have defined the density slopes $\gamma_{\mathrm{DM}}(r)$ of the DM and $\gamma_g(r)$ of the gas as
\begin{equation}
\label{a17}		
\gamma_{X} (r) \equiv \frac{ \mathrm{d} \ln {\rho}_{X}}{ \mathrm{d} \ln r }~,
\end{equation}
with $X$ standing for either $\mathrm{DM}$ or $g$. We stress that Eq.~(\ref{a16}) captures a crucial point of the present 
investigation: only the gas density slope appears on its left-hand side, whereas only quantities pertaining to the DM 
appear on its right-hand side. It should be appreciated that this result merely relies upon the equality of gas and DM 
temperatures and -- unlike in previous studies  \citep{1976A&A....49..137C,masasu, susama} -- no assumption 
is being made about the actual gas temperature structure (e.g. isothermal or polytropic).

Next, we use the fact that the DM anisotropy profile ${\beta} (r)$ turns out to be almost linearly related to the slope 
of the DM density profile $\gamma_{\mathrm{DM}}(r)$.
This result has been obtained from numerical simulations and holds with a scatter of about 0.05 \citep{hansenmoore,hansenstadel}. 
It has recently been confirmed by high-resolution numerical simulations \citep{navarro2008} and moreover it has been 
derived analytically \citep{2008arXiv0812.1048H} (see also \cite{2008ApJ...682..835Z,wojtak,2007ApJ...666..181S}).

\begin{figure}[tb]
        \centering
        \includegraphics[angle=0,width=0.49\textwidth]{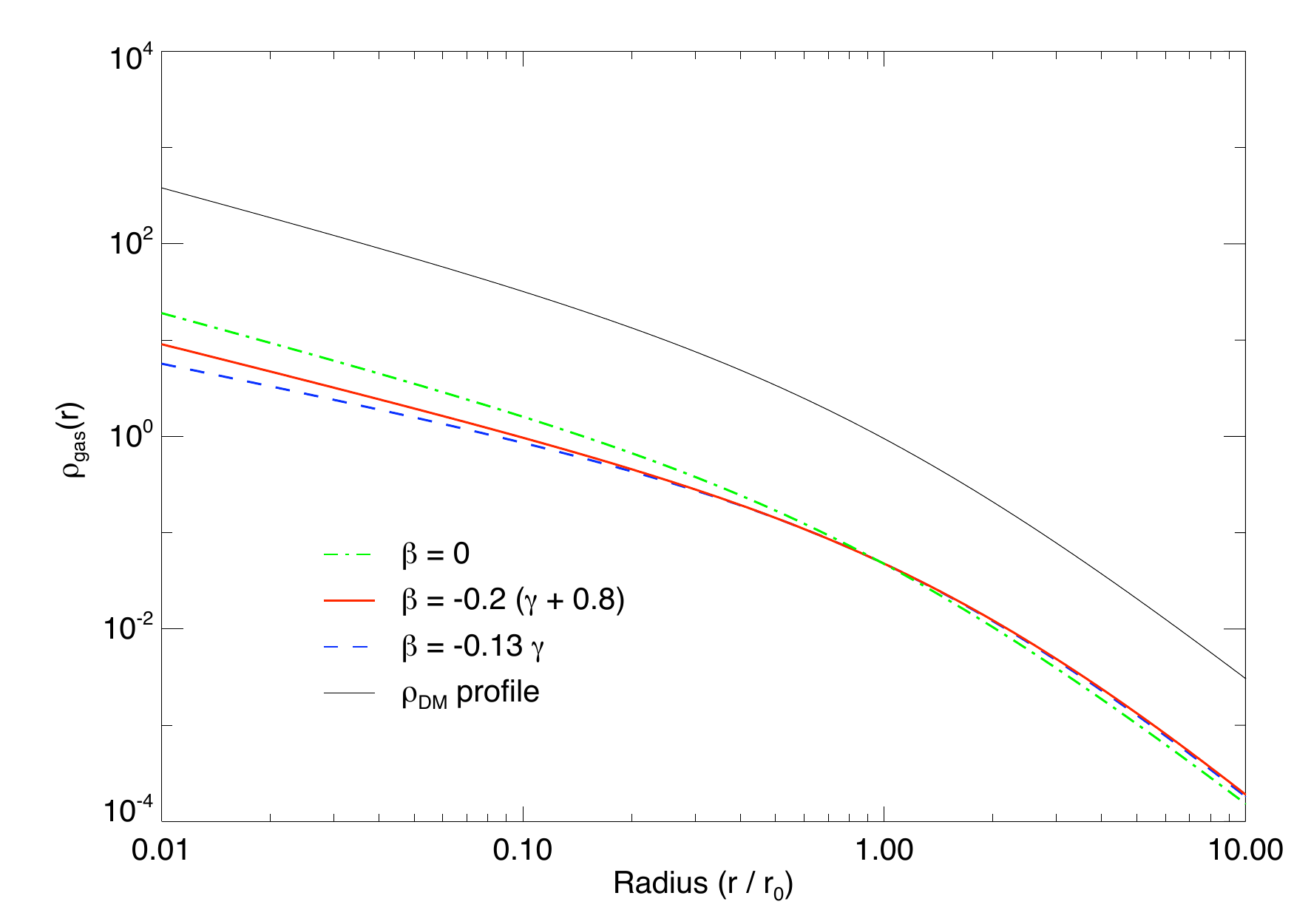}
        \caption{The derived gas density profile, assuming that
        $\rho_g / \rho_{\mathrm{DM}} = 10 \%$ at $r_0$, which is the scale length of the NFW profile.
        The upper curve (black) is the DM density, and the 3 lower lines
        show gas profiles modelled with extreme
        variations in the possible DM velocity anisotropy (green dot-dashed is
        isotropic ($\beta =0$), red solid is using $\beta = - \, 0.2 \, (\gamma + 0.8)$ \citep{hansenstadel},
        and blue dashed is using $\beta = - \, 0.13 \, \gamma$~\citep{2008arXiv0812.1048H}). }
\label{fig:nfw}
\end{figure}

\begin{figure}[htb]
        \centering
        \includegraphics[angle=0,width=0.49\textwidth]{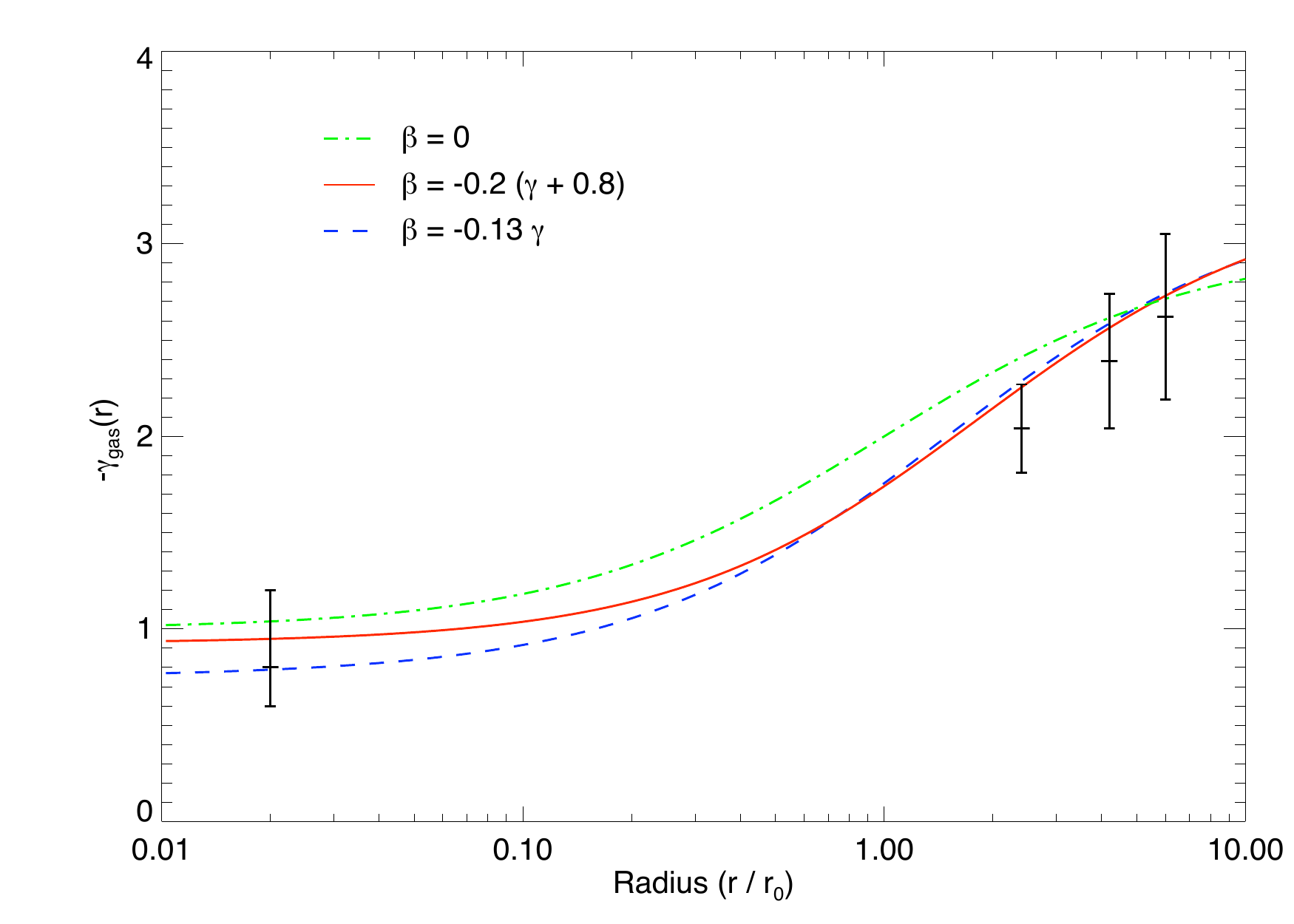}
        \caption{The derived slope of the gas density profile, assuming an NFW profile for the DM.
        Same notation as in Figure \ref{fig:nfw}. The inner point is taken from \cite{vikhlinin} and the three 
        outer points are taken from \cite{2008arXiv0811.3556E}.}
\label{fig:nfwslope}
\end{figure}

Getting the gas density profile ${\rho}_g(r)$ now involves a few simple steps. Our only input is the 
DM density profile ${\rho}_{\mathrm{DM}}(r)$, like e.g. an NFW profile. Thanks to Eq.~(\ref{a17}), we rewrite 
the Jeans equation (\ref{a5}) as
\begin{equation}
\label{a18}
r \frac{\mathrm{d} \sigma^2_r}{\mathrm{d} r} + \sigma^2_r \, \Bigl({\gamma}_{\mathrm{DM}} (r) + 2 {\beta} (r) \Bigr) 
+ \frac{G M_{\mathrm{tot}}(r)}{r} = 0~,
\end{equation}
whose solution is easily found to be
\begin{equation}
\label{a19}
\sigma^2_r (r) =  \frac{G}{B(r)} \int_r^\infty {\mathrm{d}} r' \, \frac{B(r') \, M_{\mathrm{tot}}(r')}{{r'}^{2}}~,
\end{equation}
with
\begin{equation}
\label{a20}
B(r) \equiv \rho_\mathrm{DM} (r) \, \mathrm{exp} \left\{ - \, 2 \int_r^\infty \mathrm{d} r' \, \frac{\beta(r')}{r'} \right\}.
\end{equation}
Using the relation between $\beta (r)$ and $\gamma  _{\mathrm{DM}} (r)$, we finally obtain the gas density profile from 
Eqs.~(\ref{a16}) and (\ref{a17}).  

In practice, such a procedure can be implemented iteratively. In first approximation, we assume that the gas 
mass contribution is negligible, so that we have $M_{\mathrm{tot}}(r) = M_{\mathrm{DM}}(r)$. 
In the next iterations, we include the gas mass in the calculation of $\sigma_r^2(r)$. Although the gas mass is taken into account perturbatively, any desired accuracy can be achieved by a sufficient number of iterations.

An example of the application of this strategy is shown in 
Figure \ref{fig:nfw}, where the DM density is assumed to follow an NFW profile (black solid line). The three lower lines are 
the gas density profiles obtained with a range of different possible DM velocity anisotropy profiles. The details of the gas 
density profile are easier seen in the slope, which is shown in Figure \ref{fig:nfwslope}. Note that for an inner DM slope of 
about $1$ (in agreement with the observations \citep{2006MNRAS.368..518V,pointe}) the inner {\it gas} slope should also 
be close to $1$. This is in good agreement with the fits from \cite{vikhlinin}, which have an average of $0.8$ for the extrapolated inner slope. Also the slopes found by \cite{2008arXiv0811.3556E} agrees with an NFW profile.

\begin{figure}[htb]
        \centering
        \includegraphics[angle=0,width=0.49\textwidth]{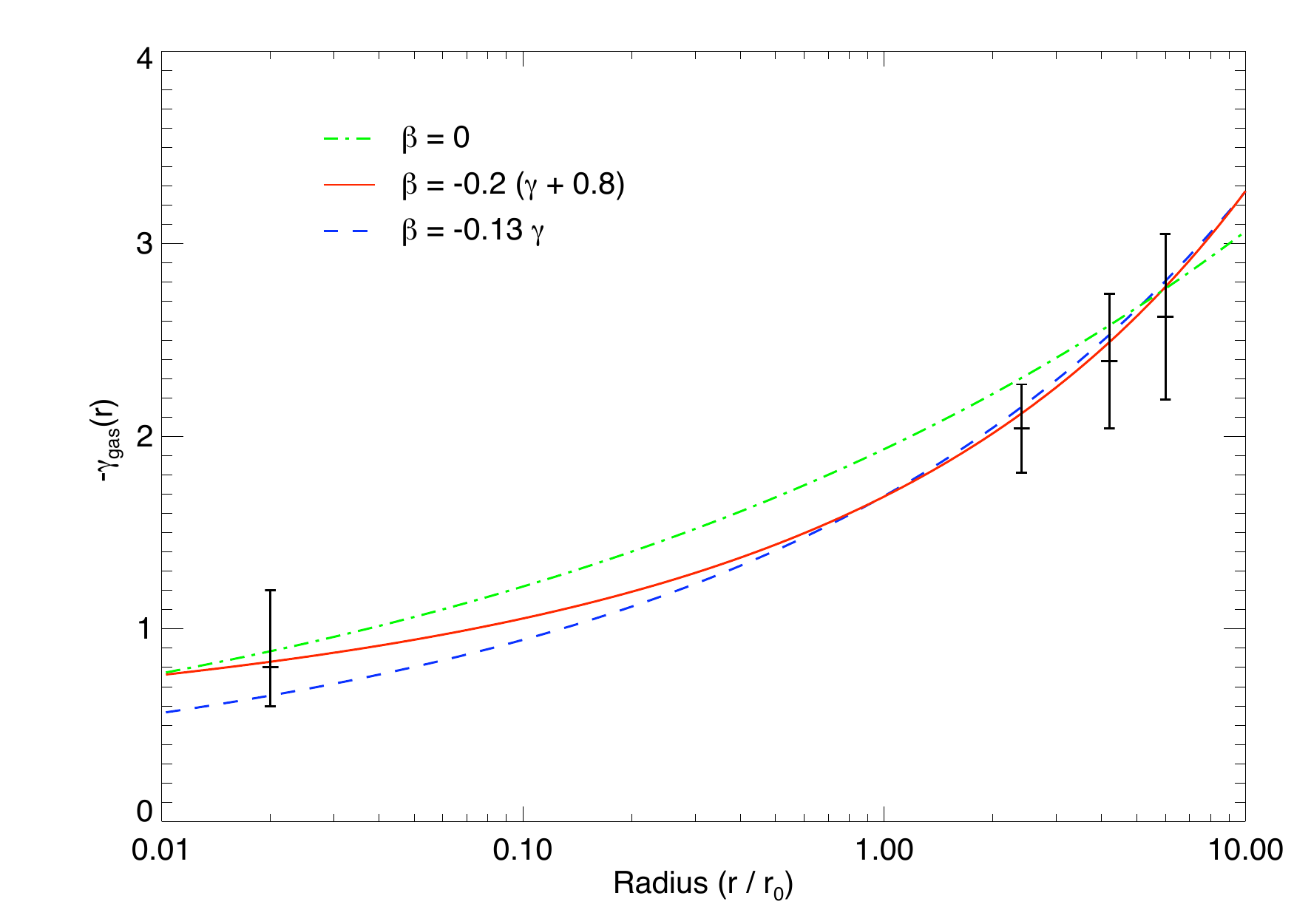}
        \caption{The slope of the gas density profile, assuming a Sersic profile with $n=5$ for the DM. 
        Same notation as in Figures  \ref{fig:nfw} and 2.}
\label{fig:sersicslope}
\end{figure}

A widely used alternative to the NFW profile is the Sersic (or Einasto) profile, which generalizes the 
de Vaucouleurs profile traditionally used to fit the optical surface brightness of elliptical galaxies. It has been shown 
that the Sersic profile models the deprojected DM density at least as well as the NFW~\citep{2004MNRAS.349.1039N,
2006AJ....132.2685M,2007ApJ...666..181S}. This profile contains 3 free parameters: two scaling constants for the 
density and the radius -- $\rho_0$ and $r_0$ respectively -- and one shape parameter $n$
\begin{equation}
\label{a23}
\rho(r) = \rho_0 \exp \left[- \, b_n \left( \left( \frac{r}{r_0} \right)^\frac{1}{n} - 1 \right) \right]~.
\end{equation}
The constant $b_n$ is a function of the index $n$ and is tabulated e.g. by \cite{mazure}. The radial velocity dispersion
$\sigma_r$ derived from the Sersic profile has, like the NFW profile, the property of reaching its maximum near 
$r_0$ where the slope is $\gamma=2$. In Figure \ref{fig:sersicslope} we present the gas density slope, assuming a 
Sersic profile for the underlying DM density. There is not sufficient statistical power in the data to discriminate 
between the underlying DM density and/or velocity anisotropy profiles from this analysis.

Both the NFW and the Sersic profile are consistent with observations because they have the appropriate slope in the inner and outer observed region. Since we cannot exclude one or the other by relying upon their shape, we choose the NFW model for the underlying DM in the rest of our treatment.

\begin{figure}[htb]
        \centering
        \includegraphics[angle=0,width=0.49\textwidth]{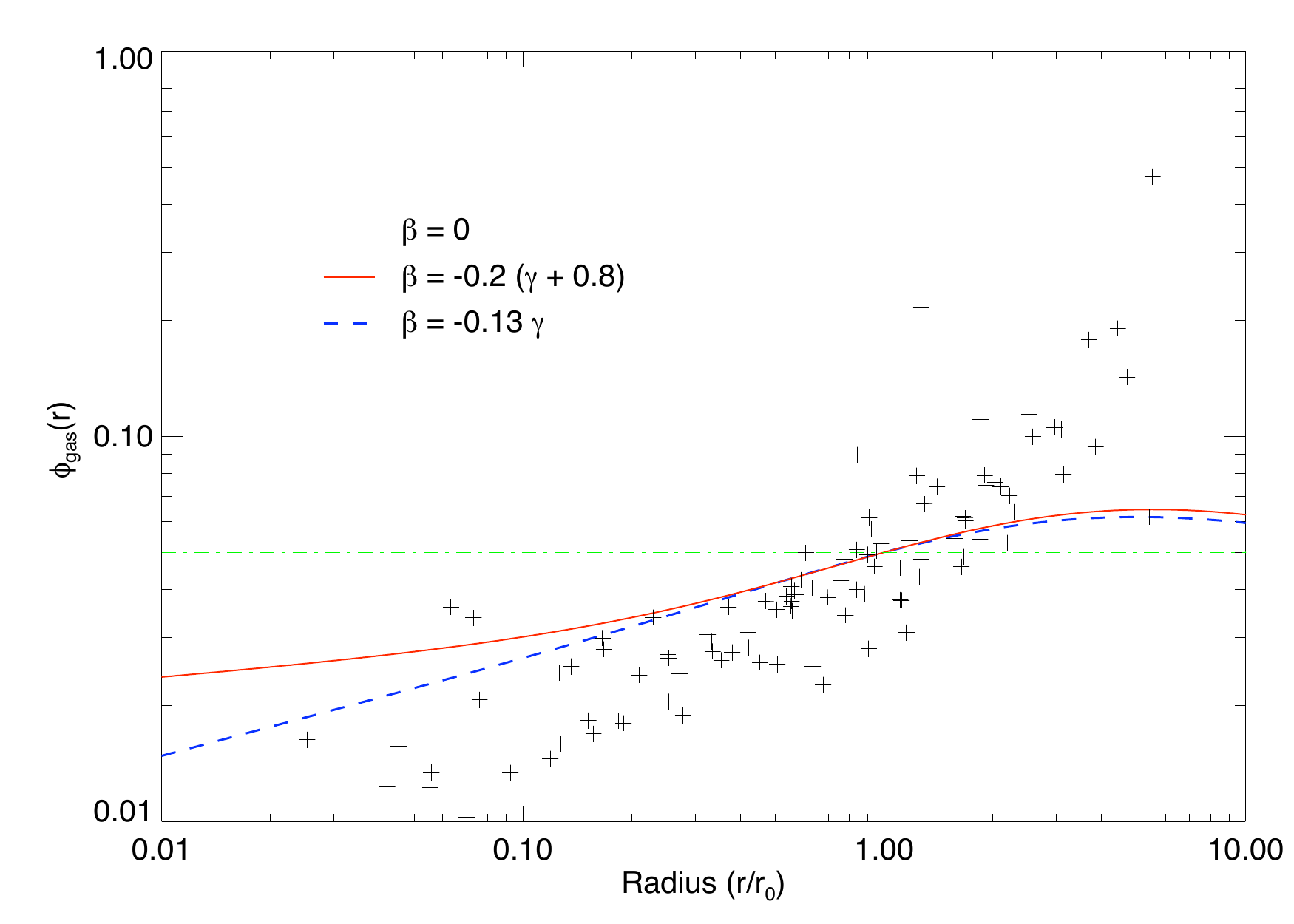}
        \caption{The observed $\phi_g$ from 16 relaxed galaxy
        clusters. We have scaled the gas mass fraction with free parameters, 
        to make the radial variation more visible.        
        There is a rough trend that the gas mass fraction  increases as
        $r^{0.5}$, which is in clear disagreement with the assumption that
        the DM velocity anisotropy should vanish. The 3 curves all assume
        an NFW density profile, and different assumed
        connections between the DM slope and velocity anisotropy, as in Figure \ref{fig:nfw}.}
\label{fig:fgas}
\end{figure}

Since the gas density profile differs from the underlying DM density profile, there will also be a radial variation in the 
local  and cumulative gas fractions, which are defined as
\begin{equation}
\label{a25}
\phi_g (r)=\frac{\rho_g(r)}{\rho_{\mathrm{tot}}(r)}
\end{equation}
and
\begin{equation}
\label{a24}
f_g (r) = \frac{M_{g}(r)}{M_{\mathrm{tot}}(r)}~,
\end{equation}
respectively.
In order to test this in more detail, we used the 16 clusters analysed in \cite{host2009}, which is a selection
of highly relaxed clusters at both low and intermediate redshifts 
\citep{2004A&A...413..415K,piffaretti2005,2007MNRAS.379..518M} 
observed with {\it XMM-Newton} and {\it Chandra}. Under the assumption of hydrostatic equilibrium, we find 
the local gas fraction exhibited in Figure \ref{fig:fgas}. The local gas fraction clearly increases as a function of radius, 
which demonstrates that the DM velocity anisotropy cannot vanish (green dot-dashed line in Figure \ref{fig:fgas}). 
The gas density fraction roughly increases as a power-law in radius and we have approximately 
$f_g (r) \sim r^{0.5}$. The solid (red) and dashed (blue) lines are for NFW DM profiles, with different radial 
DM velocity anisotropies. From Figure \ref{fig:fgas} there is a clear difference between the data and the predictions
in the outer region, which may be due to an underestimation of the total mass due to breakdown of hydrostatic 
equilibrium \citep{valpiff}.

It is important to keep in mind that these profiles do not contain any free parameters. Every quantity is calculated from the dark matter distribution alone.

The agreement in the inner region is better and it should be kept in 
mind that different DM density and velocity anisotropy profiles give rise to different curves. It may therefore
be possible to use the shape of $\phi_g(r)$ to recover these DM profiles  in the future.

As a further step, we discuss some of the implications of our main result. Indeed, with a full description of the gas that is directly derived from the dark matter distribution, we can predict additional observable quantities besides the gas fraction described above.

One of these quantities is the entropy, which is often characterized by the adiabatic coefficient $K_g$ of the gas
\begin{eqnarray}
K_g = \frac{k_B T}{\mu m_p} \rho^{-2/3} = \sigma^2_g \, \rho^{-2/3}~.
\end{eqnarray}
Our previous results entail that these profiles are almost perfect power laws
regardless of the $\beta$ profile. The slope changes slightly for the different 
$\beta$ profiles (between 1.1 and 1.3). This theoretical prediction is in good 
agreement with the X-ray observations, which generally produce power-law
entropy profiles \citep{piffaretti2005,2005A&A...429..791P,2006ApJ...643..730D}.

Another quantity that we are able to predict is the gas X-ray emissivity $\epsilon$. As a matter of fact, $\epsilon$ can be estimated either analytically -- because $\epsilon \propto n^2 T^{\frac{1}{2}}$ -- or by numerical codes like MeKaL \citep{mewe85} in order to include the line emission contribution. The latter strategy is especially well suited for cooler clusters, because a substantial amount of their luminosity stems from emission lines. On the other hand,  the luminosity of hotter clusters is dominated by the continuum emission. In either case, the surface brightness can be inferred from a given DM profile and  this 
can in turn be compared with observations. In this way, it is possible to construct an algorithm that adjusts the proposed DM profile until the surface brightness best-fits observations and thereby single out the optimal DM profile. { It is clear that this method will be limited due to clumpiness, which may be crucial in the outer parts, and one may have to combine the X-ray analysis with SZ observations to correctly extract all the profiles. We hope to discuss these aspects in more detail in the future.}

Whereas numerical simulations have demonstrated that the gas and dark matter temperatures
are equal in large parts of a galaxy cluster, they cannot probe the very centre of the clusters. It is 
therefore possible that $\kappa$ in Eq.~(\ref{a13}) departs from unity as $r \to 0$ if there is significant cooling 
or heating. However, in our derivation of the gas density profile we have assumed $\kappa =1$ everywhere.

It goes without saying that we can turn the argument around and use the observed gas profile to determine $\kappa$.
Basically, we can  insert Eq.~(\ref{a13}) 
 into Eq.~(\ref{a15}) and solve for $\kappa$. Furthermore, since we are now interested in the central region where 
$\beta$ is likely to be vanishingly small, we discard all terms involving $\beta$ in the resulting equation. So, in 
place of Eq.~(\ref{a16}) we presently get
\begin{equation}
\kappa = \frac{ \gamma_g + {\mathrm{d}} \ln \sigma^{2}_r / {\mathrm{d}} \ln r}{ \gamma_{\mathrm{DM}} + 
{\mathrm{d}} \ln \sigma^{2}_r / {\mathrm{d}} \ln r}~, 
\end{equation}
which in principle allows to measure $\kappa$ directly from X-ray observations. Such measurement
can be used or tested in future numerical 
simulations when the increased particle number will allow simulations to probe closer to the cluster centre.

\section{Conclusions}

We have shown that all properties of the hot X-ray emitting gas in galaxy clusters are completely 
determined by its underlying DM structure.  
Apart from the { standard conditions of spherical symmetry and hydrostatic equilibrium 
for the gas, our proof is based on the Jeans equation for the DM and} two simple relations which 
have recently emerged from numerical simulations. One is the equality of the gas and DM 
temperatures. The other is an almost linear relationship between the DM velocity anisotropy profile 
and its density slope. For DM distributions described by the NFW or the Sersic profiles, the resulting gas density profile, 
the gas-to-total-mass ratio profile and the entropy profile are all in good agreement 
with X-ray observations. We feel that our result is conceptually relevant in itself. Moreover, it allows to 
predict the X-ray luminosity profile of a cluster in terms of its DM content alone. Therefore, a new 
strategy becomes available to constrain the DM morphology in galaxy clusters from X-ray observations
 \citep{frederiksen09}.
This strategy may constrain morphology parameters better because of the tighter errors on surface brightness, 
but requires the structure to be very relaxed,
{ and ignores clumpiness}, and thus cannot be used on every cluster.
Our results can also be used as a practical tool for creating initial conditions for rea\-listic cosmological structures 
to be used in numerical simulations.

\noindent

\section*{Acknowledgements}
It is a pleasure to thank Andrea Morandi and Rocco Piffaretti for letting us use their reduced X-ray observations. 
One of us (M. R.) thanks the Dipartimento di Fisica Nucleare e Teorica, Universit\`a di Pavia, for support. The Dark 
Cosmology Centre is funded by the Danish National Research Foundation.

\end{document}